# Structural, electrical and magnetic properties of $Mo_{1-x}Fe_xO_2$ ($x$=0-0.05) thin films grown by pulsed laser ablation

Ram Prakash, D. M. Phase and R. J. Choudhary[a]

UGC-DAE Consortium for Scientific Research, Indore-452001, India.

Ravi Kumar

Inter University Accelerator Center, N. Delhi-110067, India.

## Abstract

We report the growth of undoped and Fe (2 and 5 at%) doped molybdenum oxide thin films on c-plane of sapphire substrate using pulsed laser ablation. X-ray diffraction results show that the films are oriented in (100) direction and have monoclinic structure based on $MoO_2$ phase as also supported by Raman spectroscopy. The x-ray photoelectron spectroscopy reveals chemical state of Fe is +2, which favors the substitutional occupancy of Fe ion in the $MoO_2$ matrix. The room temperature resistivity of all the films are very low (~100 μΩcm). The Fe doped samples show ferromagnetic behavior at room temperature.

---

[a] Corresponding author: ram@csr.ernet.in

**INTRODUCTION**

For the last several years, molybdenum oxide has attracted attentions because of their potential applications in gas sensing devices,[1-2] optically switchable coatings,[3] catalysis[4] etc. It also exhibits electrochromism, photochromism after intercalating with an apropiate cation (such as $Li^+$, $Na^+$), making suitable for use in display devices,[5] smart windows[6] and storage batteries. The demonstration of such a wide range applications is due to the non-stoichiometric nature of molybdenum oxide and to the occurrence of several different allotropes and phases of molybdenum oxide (such as $MoO_3$, $MoO_2$, $\beta$-$MoO_3$, $Mo_4O_{11}$, etc.). The crystal structure of $MoO_2$ is monoclinic while $\alpha$–$MoO_3$ and $\beta$–$MoO_3$ have orthorhombic and monoclinic structure respectively. The dependence of electrical property on oxygen concentration is such that $MoO_3$ is optically transparent[7] and electrically insulating in nature while $MoO_2$ is metallic.[8] This provides a window to optimize oxygen concentration between these two members of molybdenum oxide in a way to achieve transparent conducting semiconductor with desired band gap. Besides the above-mentioned attracting physical and optical applications, if another degree of dimensionality in terms of magnetic property could be induced in the system by doping some magnetic impurity, the resulting device will grant a boost to the existing $MoO_3$ based technology. Though the room temperature ferromagnetism in transition metal elements doped in other oxides system has been realized, there have been controversies regarding the cause of origin of ferromagnetism in the system. Some of the reports suggest it to be due to magnetic impurity clusters while some reports put in it to be charge carrier mediated ferromagnetism.[9-11] While these questions are still being contested, it would be interesting to look at the effect of Fe doping in $MoO_x$ thin films on its structural, electrical, magnetic and optical properties.

There are a few reports[1-3,7,12-17] available for growth of $MoO_x$ thin films on various substrates deposited by various deposition methods. Cardenas et al.[16] reported growth of $MoO_3$ thin films by continuous wave $CO_2$ laser–assisted evaporation and their optical characterization suggested the formation of oxygen vacancies at higher substrate temperatures. Bhosle et al.[14] reported the epitaxial growth and properties of $MoO_x$ (2<x< 2.75) films by pulsed laser ablation and studied the effect of thermal annealing on these films. Choi et al.[21] performed surface properties using X-ray photoelectron spectroscopy (XPS) of as prepared and reduced molybdenum oxide and they show that $MoO_3$ reduced to $MoO_2$ by $H_2$ at 400 °C. In the present paper we report the growth of Fe (0-5 at %) doped $MoO_2$ thin films on c-plane sapphire substrates by pulsed laser ablation and investigated their structural, electronic, electrical and magnetic properties. This is the first report on Fe doped molybdenum oxide thin films in best of our knowledge.

**EXPERIMENTAL**

The thin films of undoped and Fe (0-5 at %) doped $MoO_2$ were grown on c-plane sapphire single crystal substrates by pulsed laser ablation technique. Molybdenum oxide ($MoO_3$) target, used for deposition of undoped $MoO_2$ films, was prepared by sintering the pallet of Molybdenum oxide powder at 750 °C for 12 hours. The Fe doped targets were prepared by standard solid-state route from α-$Fe_2O_3$ and $MoO_3$ powders. These targets were sintered at same temperature as used for undoped target. The single phase and composition of bulk target was examined by x-ray diffraction (XRD) and energy dispersive analysis of x-rays (EDAX) techniques respectively. After preparing the target we have deposited thin films of all targets on c-plane sapphire substrates. A pulsed KrF (wavelength = 248 nm) Excimer laser was used for ablation. The energy density of the laser beam was kept at 3 J/cm$^2$ with repetition rate of 10 Hz. The chamber was evacuated to base pressure of $1\times10^{-6}$ Torr and deposition was carried out at $1\times10^{-4}$ Torr of oxygen background pressure. The deposition was performed at the temperature 550 °C for 20 min. The thicknesses of all the films are around 200 nm as measured by stylus profilometer. The X-ray diffraction (XRD) of the films was carried out using Rigaku diffractometer with CuKα radiation (λ=1.54 Å). The X-ray photoelectron spectroscopy (XPS) measurements was performed using Omicron energy analyzer (EA 125) instrument with Al Kα (1486.6 eV) x-ray source. The pressure of the analyzer chamber was in order of $10^{-10}$ Torr during the XPS measurement. The analysis of different oxidation states of ions was performed by deconvolution of unresolved peaks. The spectrometer was calibrated using Au-4f$_{7/2}$ core level (at binding energy =84 eV) spectra. The values corresponding to C 1s peak were used as a reference for spectrum analysis. The survey scan spectra for all samples were recorded at 50 eV pass energy while in case of narrow scans the pass energy was kept at 40 eV. For study of the vibrational properties of these films, laser Raman spectroscopy was employed. Raman spectra were recorded at room temperature using Micro Raman spectrometer (Model HR-800, Jobin Yvon) employing He-Ne laser (λ=632.8nm). The measured resolution of spectrometer is 1 cm$^{-1}$.

Spectra were collected in backscattering geometry using charge-coupled device (CCD) with laser power 9 mW and incident laser power focused in a diameter of 2μm. Notch filter is used to suppress the Reighley light. The electrical resistivity measurement was performed using standard four probe technique. Magnetization measurements were performed using vibrating sample magnetometer (VSM) technique (Lakeshore, Model 7401) at room temperature.

**RESULTS AND DISCUSSION**

Figure 1 shows the XRD pattern of undoped, 2 and 5 at% Fe doped $MoO_2$ thin films measured in θ-2θ geometry. Film peaks are marked as *F*. The peak positions of the patterns match well with the XRD pattern of monoclinic $MoO_2$ phase (JCPDS No. 78-1069). We also notice that all the films are single phase and highly a-axis oriented, as earlier reported by Bhosle et al.[14] We do not observe the presence of any other phase of molybdenum oxide or iron oxide in either undoped or Fe doped films. Inset in Figure 1 shows the variation of out of plane lattice parameter, as calculated from the XRD pattern. It is observed that there is no considerable change in the lattice parameter after Fe doping. Since XPS data suggests that Fe is in +2 valence state (as will be discussed later section), we note that the ionic radii of $Fe^{+2}$ is closer to $Mo^{+4}$ state, hence the doping of Fe into the $MoO_2$ matrix would be most favorable when Fe is in +2 valence state. The (200) peak of each samples were fitted to get peak position and width, which was used to calculate lattice parameter and particle sizes. We have calculated the grain size (D) of the film using Debye - Scherrer formula[18] given by

$$D = 0.94 \times \lambda / (B Cos\theta).$$

Where, λ is the wavelength of the x-ray source and B is the full width at half maxima (FWHM) of individual peak at 2θ (where θ is Bragg angle). The calculated lattice parameter and particle size are listed in the table 1. There is no observable change in lattice parameter of the film but particle size change slightly in 2 at% Fe doping while change drastically after 5 at% Fe doping. Particle size for undoped film is 30 nm while it deceases to 17 nm for 5 at% Fe doped film.

This decrease in the particle size is related to defects produced in film due to iron doping.

To investigate chemical state and surface composition of elements present in the samples we have performed XPS measurement for undoped and Fe doped $MoO_2$ thin films deposited on sapphire substrate by pulsed laser ablation. Figure 2 shows the XPS survey spectra of the all three films sample. As molybdenum oxide is very sensitive to environment therefore we have sputtered the films using 1 keV $Ar^+$ ion for 5 min duration to avoid the surface contamination. These survey spectra clearly show that in films have only Mo, O and Fe in doped case are present in the spectrum. No other impurity was present in the samples except signal from the elemental Cu which comes from sample holder. All peaks present the spectrum were indicated in the figure at appropriate point. To further analysis core level spectra of the samples we have performed the detailed scan for Mo-3d, and O-1s core levels for all samples while Fe-2p core level for iron doped samples.

Since it is realized that Mo has various oxidation states, therefore, in order to identify the actual oxidation state of Mo, we performed the Mo-3d core level spectra. In figure 3 we show the Mo-3d core level spectra for undoped and Fe doped samples. The observed Mo3d core level spectra are characteristic of the complex mixture of Mo $3d_{5/2}$, $_{3/2}$ unresolved spin orbit doublet of molybdenum in its oxidation states of +4, +5 and +6. These spectra were fitted using a non-linear fitting program with three-spin orbit doublet component, which correspond, to three different oxidation state of Mo. The spin orbit splitting was kept fixed at 3.17 eV for all three components of Mo3$d_{3/2}$ and Mo3$d_{5/2}$ peaks. The intensity ratio for spin orbit doublet Mo3$d_{5/2}$ and Mo3d$3_{/2}$ was kept at 1.5. In case of undoped MoO$_2$ thin film, the Mo3$d_{3/2}$ peaks with binding energy values at 228.9, 230.2 and 233.3 eV can be assigned to Mo$^{+4}$, Mo$^{+5}$ and Mo$^{+6}$ states respectively, which match with the previous reported[15,19-22] values. The relative concentration of each species (Mo$^{+4}$, Mo$^{+5,}$ and Mo$^{+6}$) was determined by calculating area under the resolved peaks present in the film. After calculating the area under the convoluted peaks, we note that Mo$^{+4}$, Mo$^{+5}$ and Mo$^{+6}$ states are 61, 34 and 5% respectively. To study the effect of Fe doping on MoO$_2$ we have fitted the Mo3d spectra for Fe doped samples also. The fitting results are listed in table 2. We note that as we dope Fe into the system, Mo$^{+4}$ concentration increases at the cost of Mo$^{+5}$ states. We observed a tail at higher binging energy side in Mo-3d line shape as observed by M. Sing et al.[22] in case of metallic K$_{0.3}$MoO$_3$ suggesting that our films are mostly contained by metallic MoO$_2$. Werthein et al.[23] concluded from detailed analysis of Mo-3d spectrum that line shape is caused by different screening channels in the photoemission final state; one in which core hole is screened by charge transfer from valence band to empty Mo 4d conduction band states (as in case of insulating MoO$_3$), and second in which electron already in conduction band (as in case of metallic MoO$_2$) performed metallic screening. The second process, in a final state of lower binding energy, is responsible for the observed tail typical of the collective response of conduction electrons in metals.

Since MoO$_2$ is very sensitive to the atmospheric oxygen, on film's surface one would expect adsorption of oxygen, which would have different chemical environment than the oxygen attached to Mo atoms. Therefore, to investigate the chemical state of oxygen ion present in the films we have recorded O1s core level spectra for undoped and doped MoO$_2$ samples. The O1s spectra for undoped and Fe doped samples are shown Fig. 4. It is clear from these spectra that most of the oxygen ions in all samples are bonded with Mo ions. The binding energy peaks positions for O1s bonded with Mo are 530.5 eV which matches well with O$^{-2}$ ionic states as reported values in the literature.[24] In all the films, a small amount of oxygen (around 10%) at binding energy positions around 532.5 eV is related with water adsorbed at sample surface as reported in the literature.[25]

Finally, to investigate Fe chemical state in the Fe doped samples we have performed detailed spectra of Fe2p core level for Fe doped thin films, which are shown in figure 5. From the Fe2p spectra it can be seen that Fe-2p$_{3/2}$ and Fe-2p$_{1/2}$ binding energy positions for 2% and 5% Fe doped samples are at around 709 eV and 722 eV, which are very close to Fe$^{+2}$ states as in case of FeO (709.3 for Fe-2p$_{3/2}$ and 722.3 for Fe-2p$_{1/2}$ )[26]. However, from litlerature[26] of XPS we also get that the peaks of Fe 2p$_{3/2}$ and Fe 2p$_{1/2}$ for Fe metal is located at 706.75 eV and 719.95 eV, and for Fe$^{+3}$ in Fe$_2$O$_3$ at 710.70 eV and 724.30 eV, respectively. The observed Fe2p peak positions match with the earlier reported literature[26] values for Fe$^{2+}$ states. Clearly, XPS measurements exclude the possibility of presence of metallic iron or iron cluster in the Fe doped samples.

XRD and XPS studies revealed the crystal structure and surface composition of the films and confirmed it to MoO$_2$ phase of molybdenum oxide. However, to understand the structure and composition deep inside the films we have performed laser Raman spectroscopy measurements for all three samples at room temperature with He-Ne laser source of wavelength 632.8 nm. Figure 6 shows Raman spectra of all samples recorded in the range of 100 to 1100 cm$^{-1}$ along with sapphire substrate. The Raman spectra for all samples reveal $MoO_2$ phase. The Raman mode of monoclinic $MoO_2$ were reported[27] at 207, 230, 347, 364, 458, 496, 569, 585 and 740 cm$^{-1}$ positions in which the Raman modes at 585 and 744 cm-1 were attributed to stretching vibrations of the Mo-O(I) and Mo-O(II) group in the lattice. In other literature, Raman modes are reported by L. Kumari et al.[28] for bulk $MoO_2$ at 202.9, 226.3, 346.2, 360.2, 458.2, 494.1, 567.2, 584.4 and 738.5 cm$^{-1}$. We observe Raman modes for undoped $MoO_2$ films at positions 203, 229, 358, 458, 493, 562, 736 cm$^{-1}$. These modes match with the earlier reported literature values of bulk $MoO_2$ Raman modes. The small shift in Raman modes positions of our films may be attributed to the presence of strain in the films because thin films always have substrate induced strain and growth defects. The Raman modes for 2% Fe doped thin film sample are observed at 203, 229, 347, 361, 458, 468, 494, 565, 738 cm$^{-1}$ positions. We note that additional modes at 347 and 468 cm$^{-1}$ appear after Fe doping in $MoO_2$. Though the mode at 347 cm$^{-1}$ is attributed to the $MoO_2$, the mode at 468 cm$^{-1}$ is not known to appear in $MoO_2$. Also this mode can not be attributed to any other phase of either molybdenum oxide or iron oxide. A possible reason for this additional mode could be that if Fe replaces Mo in the crystal structure, one would expect a different coupling strength between Fe and neighboring atoms as compared to that at Mo site, though the arrangement of oxygen atoms would remain same for both the sites. The appearance of new modes after Fe doping is also reported to occur in systems like Fe doped $SnO_2$, Fe doped $La_{0.7}Sr_{0.3}MnO_3$ samples.[29,30] In these reports it is suggested that the appearance of new mode after Fe doping could be attributed to either otherwise forbidden Raman mode that become active due to symmetry breaking effect or to the phonon density of states features, which arise due to a disorder induced phenomenon. As we increase the Fe doping, the Raman modes for 5% Fe doped thin film sample are observed at 203, 229, 347, 361, 458, 468, 494, 566 and 738 cm$^{-1}$ position. These Raman modes are very close as observed in case of 2 at% Fe doped sample. These results clearly revel that the crystal structures of films are remain same after 5 at% of Fe doping.

After ensuring the structure and chemical states of ions in the films, we performed the temperature dependent electrical resistivity measurements using standard four-probe technique. The resistivity behavior with temperature is shown in the Fig. 7. This shows metal to insulator transition at temperature 101K for undoped sample while transition temperature lower to 84 K and 57 K for 2% Fe doping and 5% Fe doping respectively. We have observed very low resistivity for all the samples (~100 μΩcm). The metal to insulator transition could be due to non stoichiometric nature of the films as indicated in the XPS studies for Mo mixed oxidation states. It is recalled here that $MoO_2$ has metallic behavior while $MoO_3$ has insulating nature. Since after Fe doping, $Mo^{+4}$ concentration increases, films show more metallic behavior and hence decrease in the transition temperature.

After studying the structural, electronic and electrical resistivity behavior we performed the magnetization measurement to investigate the effect of Fe doping. The room temperature magnetization vs magnetic field hysteresis loop is shown in the Fig. 8. This shows that Fe doped films are ferromagnetic at room temperature and this magnetic moment is due the presence of doped iron in the film. The magnetic moment was calculated for 2% and 5 % Fe doping and was found to be 3.1 and 1.2 µB per Fe ion respectively. Fe being in +2 valence state (as shown by XPS measurements), spin only magnetic moment should be close to 4 $\mu_B$/Fe ion. It is recalled here that in such a system, magnetism is of itinerant nature and different models, such as RKKY, bound magnetic polaron etc. have been proposed to explain the magnetic behavior, where a decrease in magnetization with increase in magnetic impurity is observed in various DMS systems.[10,31,32] This is in consistent with our observation of magnetic measurements. On the basis of above results, it is possible that these films can be explored as conducting magnetic oxide materials.

**CONCLUSION**

We have successfully deposited undoped and Fe (2-5 at%) doped $MoO_2$ thin films on c-plane sapphire substrates by pulsed laser ablation technique. The XRD patterns show that these films are single phase and (100) oriented. XRD patterns do not change much with Fe doping suggesting that Fe replaces the Mo atoms without affecting the crystal structure of molybdenum oxide. XPS results clearly indicate that in these films Mo has mixed valence stats but $Mo^{+4}$ is in higher concentration. The Fe-2p core level spectra suggest that iron present in +2 valence state. The films reveal very low resistivity at room temperature. The magnetic measurements reveal room temperature ferromagnetic behavior in Fe doped molybdenum oxide thin films.

**ACKNOWLEDGEMENTS**

The authors are thankful to Dr. P. Chaddah and Prof. A. Gupta for the encouragement. We are also thankful to Dr. V. Sathe and Dr. N. Lakshmi for providing Raman and VSM measurements respectively. One of us (RP) is thankful to CSIR, New Delhi, India for financial support as fellowship.

REFERENCES

1. J. Okumu, F. Koerfer, C. Salinga, M. Wutting, J. Appl. Phys. **95**, 7632(2005).

2. S. H. Mohamed, O. Kappertz, J.M. Nagaruiya, T. P.L. Pedersen, R. Drese, and M. Wutting, Thin solid Films **429**, 135 (2003).

3. J. Okumu, F. Koerfer, C. Salinga, T. P. Pedersen, M. Wutting, Thin Solid Films **515**, 1327 (2006).

4. W. Zhang, A. Desikan, and S. T. Oyama, J. Phys. Chem. **99**, 14468 (1995).

5. R. J. Colton, A. M. Guzman, and J.W. Rabalais, J. Appl. Phys. **49**, 409 (1978).

6. J. N. Huiberts, R. Griessen, J. H. Rector, R. J. Wijngaarden, J. P. Dekker, D. G. DeGroot, and N. J. Koeman, Nature (London) **380**, 231 (1996).

7. C. Jullien, A. Khelfa, O.M. Hussain and G. A. Nazri, J. Crystal Growth, **156**, 235 (1995).

8. M. A. K. L. Dissnayake, and L.L. Chase, Phys. Rev. B **18**, 6872 (1978).

9. S. R. Shinde, S. B. Ogale, J. S. Higgins, H. Zheng, A. J. Millis, V. N. Kulkarni, R. Ramesh, R. L. Greene, and T. Vanketation, Phys. Rev. Lett. **92**, 166601 (2003).

10. S. B. Ogale, R. J. Choudhary, J. P. Buban, S. E. Lofland, S. R. Shinde, S. N. Kale, V. N. Kulkarni, J. Higgins, C. Lanci, J. R. Simpson, N. D. Browning, S. Das Sarma, H. D. Drew, R. L. Greene, and T. Venkatesan, Phys. Rev. Lett. **91,** 077205 (2003).

11. Ravi Kumar, Fouran Singh, Basavaraj Angadi, Ji-Won Choi, Won-Kook Choi, K. Jeong, Jong-Han Song, M. Wasi Khan, J. P. Srivastava, Ajay Kumar and R. P. Tandon, J. Appl. Phys., **100**, 113708 (2006) and reference therein.

12. N. Miyata, S. Akiyoshi, J. Appl. Phys. **58**, 1651 (1985).

13. R. S. Patil, M. V. Upane, P. S. Patil, Appl. Surf. Sci. **252**, 8050 (2000).

14. V. Bhosle, A. Tiwari, and J. Narayan, J. Appl.Phys. **97**, 083539 (2005).

15. Y. V. PLyuto, I. V. Babich, I. V. Plyuto, A.D. Van Langeveld, J.A. Moulijn, Appl. Surf. Sci. **119**, 11 (1997).

16. R. Cardenas, J. Torres, J. E. Alfonso, Thin solid films **478**, 146 (2005).

17. A. Donnadieu, D. Davazoglou, and A. Abdellaoui, Thin Solid Films **164,** 333 (1988).

18. B. D. Cullity, "Elements of X-Ray Diffraction", Addison-Wesley, Reading, MA, (1972).

19. A. Fujimori, M. Sekita, and H. Wada, Phys. Rev. **B 33**, 6652 (1986).

20. T. H. Fleisch, G. J. Mains, J. Chem. Phys. **76**, 780 (1980).

21. J. -G. Choi, L.T. Thompson, Appl. Surf. Sci. **93**, 143 (1996).

22. M. Sing, R. Neudert, H. von Lips, M.S. Golden, M. Knupfer, J. Fink, R. Claessen, J. Mucke, H. Schmit, S. Hufner, B. Lommel, W. Abmus, Ch.Jung, and C. Hellwig, Phys. Rev. **B 60**, 8559 (1999).

23. G. K. Wertheim, L. F. Schneemeyer, and D. N. E. Buchanan, Phys. Rev. **B 32**, 3568 (1985).

24. F. Werfet and E. Minni, J. Phys. C: Solid State Phys., **16**, 6019 (1983).

25. Y. L. Leung, P. C. Wong, K. A. R. Mitchell, K. J. Smith, Appl. Surf. Sci. **136**, 147 (1998).

26. A. J. Chen, X. M. Wu, Z. D. Sha, L. J. Zhauge, and Y. D. Meng, J. Phys. D: Appl. Phys., **39**, 4762 ( 2006).

27. M. Dieterle, and G. Mestl, Phys. Chem. Chem. Phys. **4**, 822 (2002).

28. L. Kumari, Y. R. Ma, C-C. Tsai, Y-W Lin, S. Y. Wu, K-W. Cheng, and Y. Liou, Nanotechnology, **18**, 115717 (2007).

29. Xavier Mathew, J. P. Enriquez , C. Mejía-García, G. Contreras-Puente, M. A. Cortes-Jacome, J. A. Toledo Antonio, J. Hays and A. Punnoose, J. Appl. Phys. **100**, 073709 (2006).

30. A.G. Souza Filho, J. L. B. Faria, I. Guedes, J. M. Sasaki, P.T. C. Freire, V. N. Freire, J. M. Filho, M. M. Xavier, Jr., F .A. O. Cabral, J. H. de Araujo, and J. A. P. da Costa, Phys. Rev. **B 67**, 052405 (2003).

31. J. M. D. Coey, M. Venkatesan and C. B. Fitzgerald, Nature Materials, **4**, 173 (2005).

32. N. H. Hong, W. Prellier, J. Sakai, A. Hassini, Appl. Phys. Lett. **84**, 2850 (2004).

**Figure caption**

**Figure 1**: X-ray diffraction pattern of undoped and Fe ( 2 and 5 at% ) doped $MoO_2$ thin films deposited on c-plane sapphire substrate by pulsed laser ablation, inset shows the variation of lattice parameter *a* with Fe doping concentration..

**Figure 2**: Survey scans of X-ray photoelectron spectra for undoped and Fe (2 and 5 at% ) doped $MoO_2$ thin films deposited on c-plane sapphire substrate by pulsed laser ablation. All peaks are marked at upper panel of figure.

**Figure 3**: XPS Mo 3d core level spectra of undoped and Fe ( 2 and 5 at% ) doped $MoO_2$ thin films deposited on c-plane sapphire substrate by pulsed laser ablation. The circle show experimental data and solid line fitted curve while peak1, 2 and 3 show $Mo^{+4}$ $Mo^{+5}$ and $Mo^{+6}$ states contribution in convoluted spectrum respectively.

**Figure 4**: XPS O1s core level spectra of undoped and Fe ( 2 and 5 at% ) doped $MoO_2$ thin films deposited on c-plane sapphire substrate by pulsed laser ablation. The circle show experimental data and solid line fitted curve while peak 1, and 2 show O1s bonded to Mo and absorbed water contribution in convoluted spectrum respectively.

**Figure 5**: XPS Fe 2p core level spectra of Fe ( 2 and 5 at% ) doped $MoO_2$ thin films deposited on c-plane sapphire substrate by pulsed laser ablation. The circle show experimental data and solid line fitted curve.

**Figure 6:** Raman spectra of undoped and Fe (2 and 5 at% ) doped $MoO_2$ thin films deposited on c-plane sapphire substrate by pulsed laser ablation. The lower panel of figure shows the Raman spectrum of c-plane sapphire substrate, which indicates that films spectra do not contain the contribution of substrate.

**Figure 7:** Plot of resistivity as a function of temperature for undoped and Fe (2 and 5 at% ) doped $MoO_2$ thin films deposited on c-plane sapphire substrate by pulsed laser ablation.

**Figure 8**: Magnetic hysteresis loop of Fe ( 2 and 5 at% ) doped $MoO_2$ thin films deposited on c-plane sapphire substrate by pulsed laser ablation.

**Table: I** XRD (200) peak position, out of plane lattice parameter, particle size and resistivity minima temperature position for undoped and Fe (2 and 5 at%) doped samples.

| Sample | Peak position (200) | Lattice parameter (nm) | Particle size (nm) | Resistivity minima temperature (K) |
|---|---|---|---|---|
| $MoO_2/Al_2O_3$ | 37.26 | 0.4826 | 31 | 101 |
| $Mo_{0.98}Fe_{0.02}O_2/Al_2O_3$ | 37.38 | 0.4810 | 27 | 84 |
| $Mo_{0.95}Fe_{0.05}O_2/Al_2O_3$ | 37.38 | 0.4810 | 17 | 57 |

**Table II**. XPS peak positions obtained by fitting of peaks for Mo3d, O1s and Fe 2p core level spectra for undoped and Fe (2 and 5 at%) doped samples the quantities in the bracket is the percentage area of particular component.

| Sample | Mo 3d$_{5/2}$ (eV) | O 1s (eV) | Fe 2p$_{3/2}$ (eV) |
|---|---|---|---|
| MoO$_2$/Al$_2$O$_3$ | 228.9 (61%) | 530.5 (89%) | |
| | 230.7 (34%) | 532.7 (10%) | |
| | 233.3 (5%) | | |
| Mo$_{0.98}$Fe$_{0.02}$O$_2$/Al$_2$O$_3$ | 228.9 (69%) | 530.5 (82%) | 709.2 |
| | 230.8 (24%) | 532.1 (12%) | |
| | 233.2 (7%) | | |
| Mo$_{0.95}$Fe$_{0.05}$O$_2$/Al$_2$O$_3$ | 228.8 (68%) | 530.5 (88%) | 709.5 |
| | 230.8 (26%) | 532.3 (11%) | |
| | 233.3 (6%) | | |

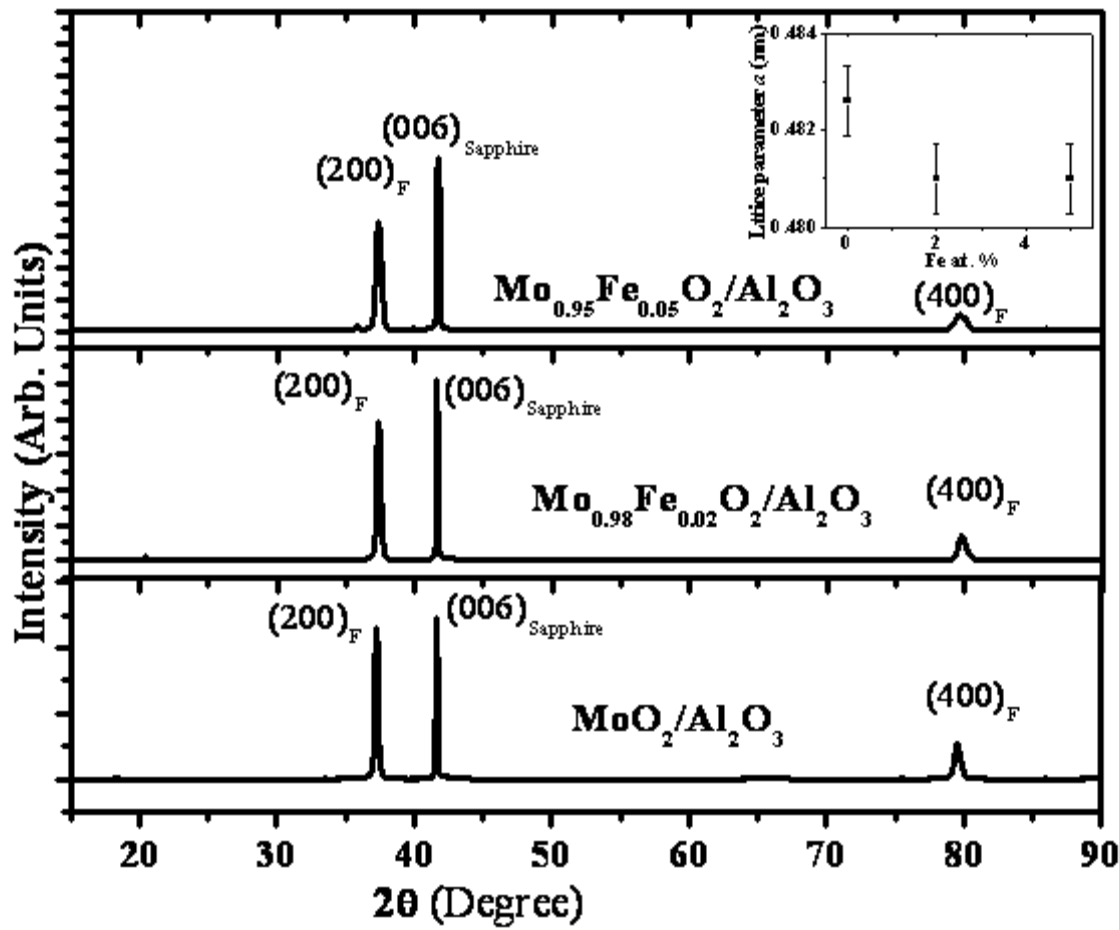

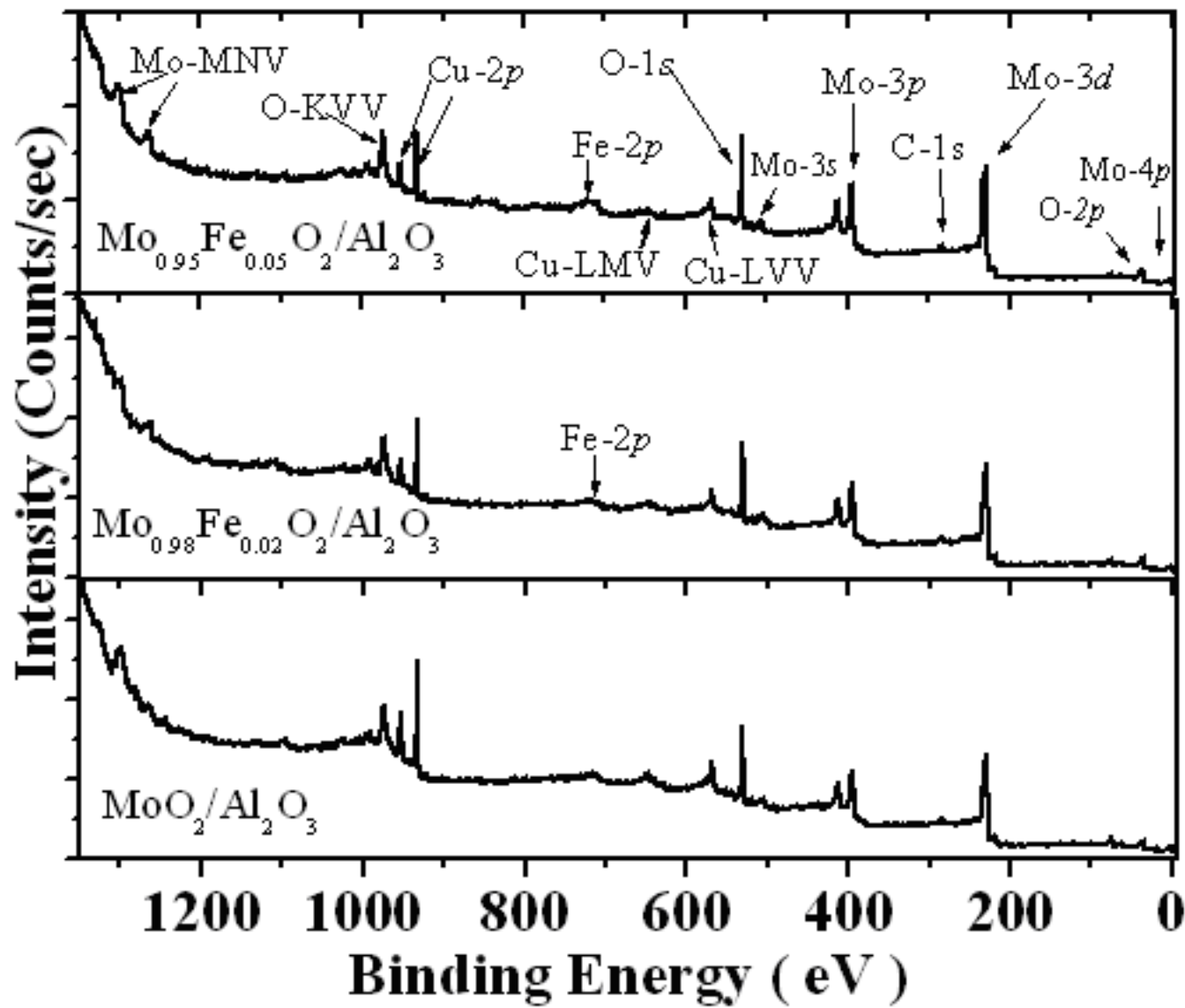

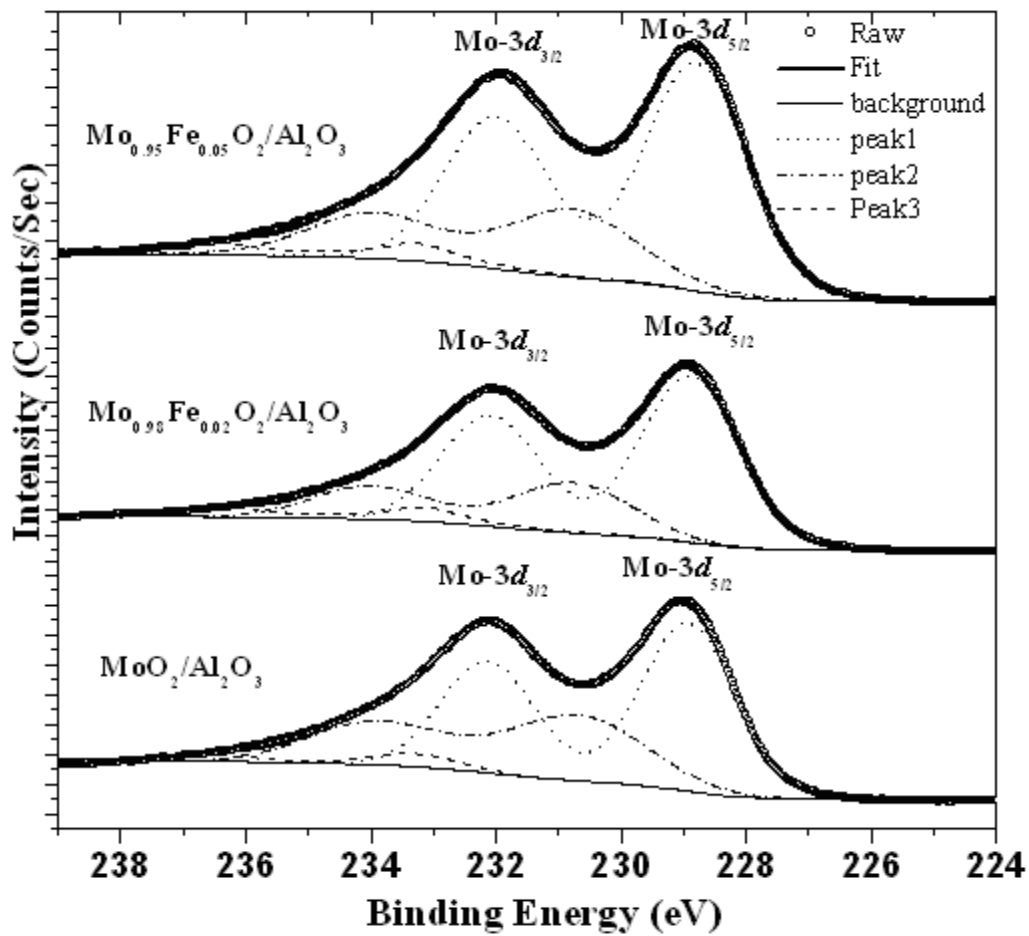

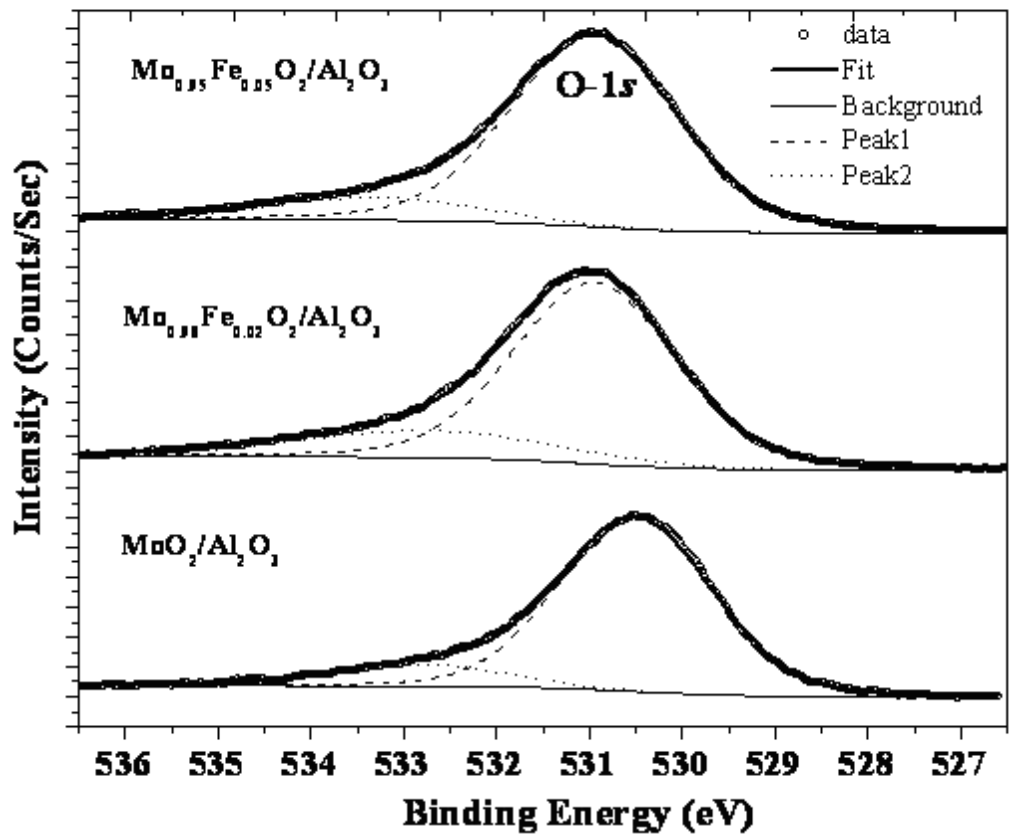

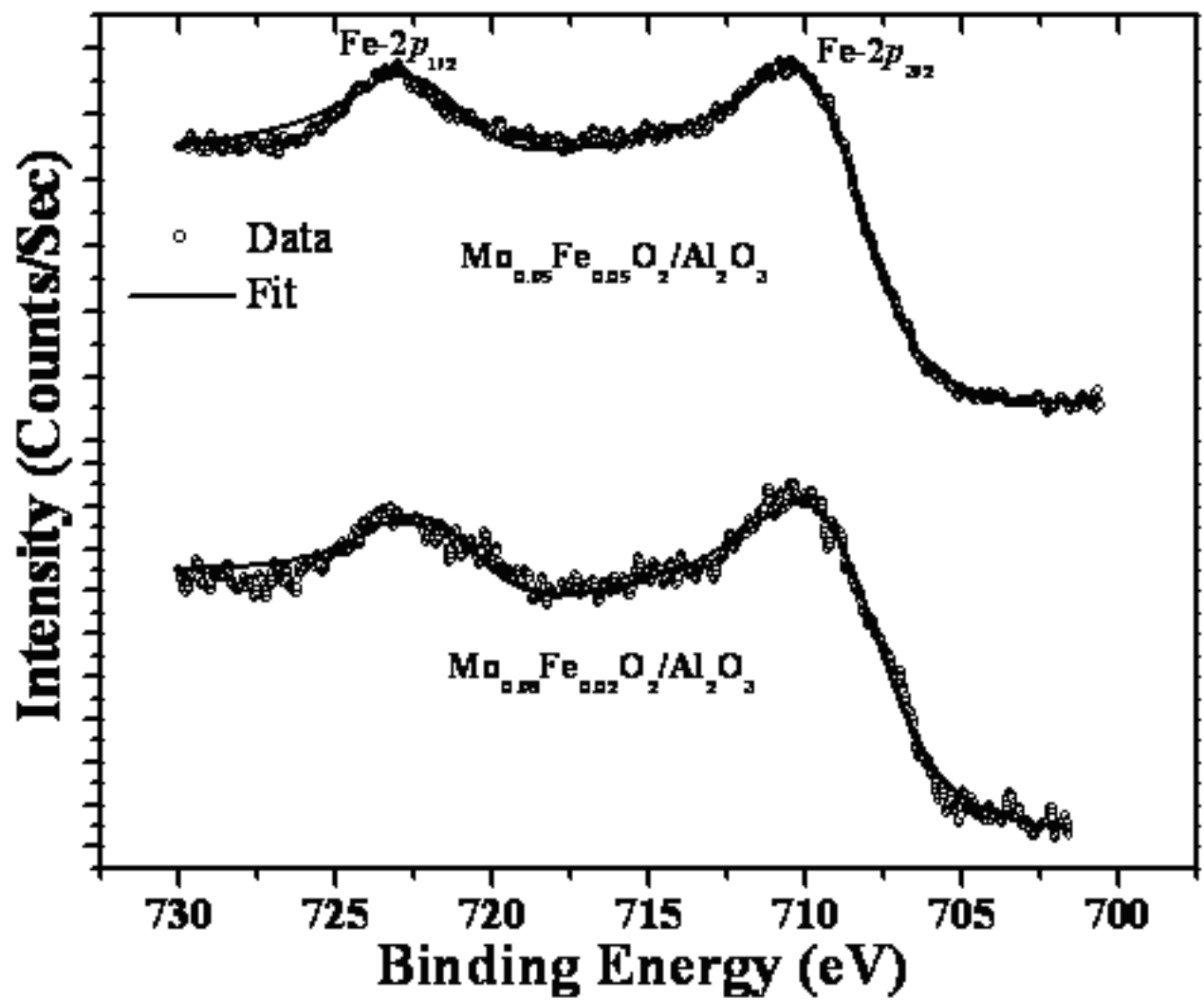

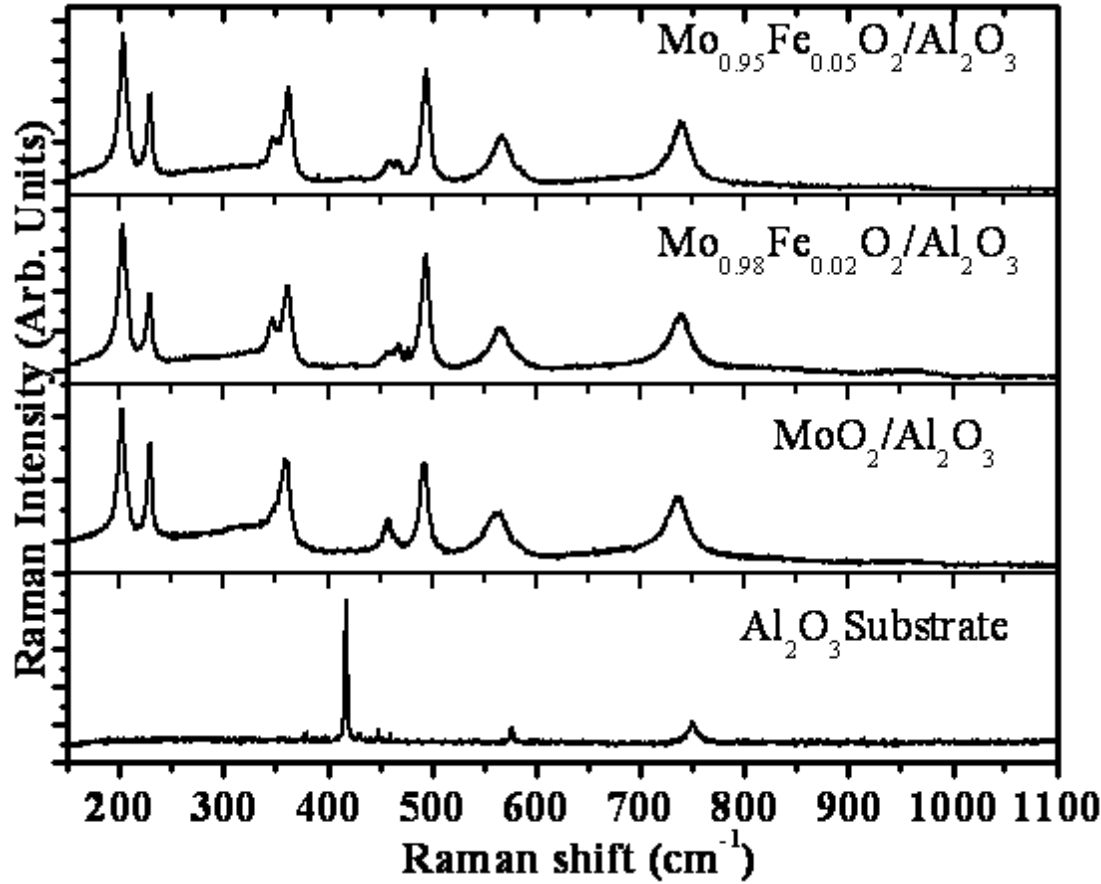

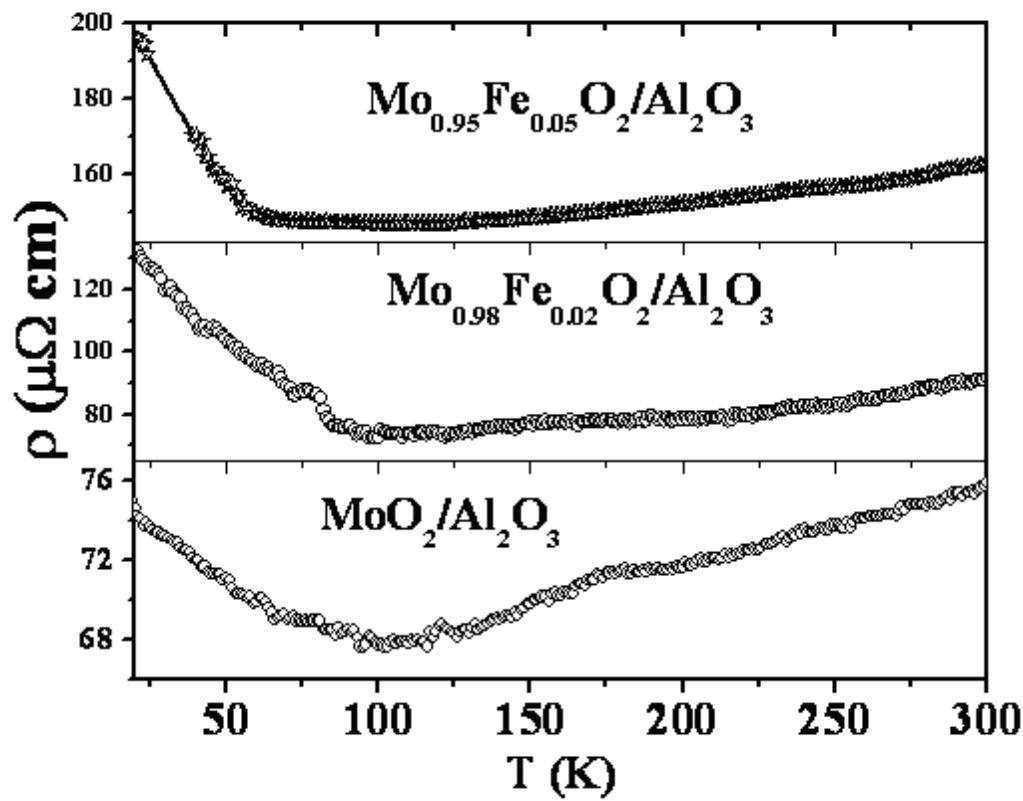

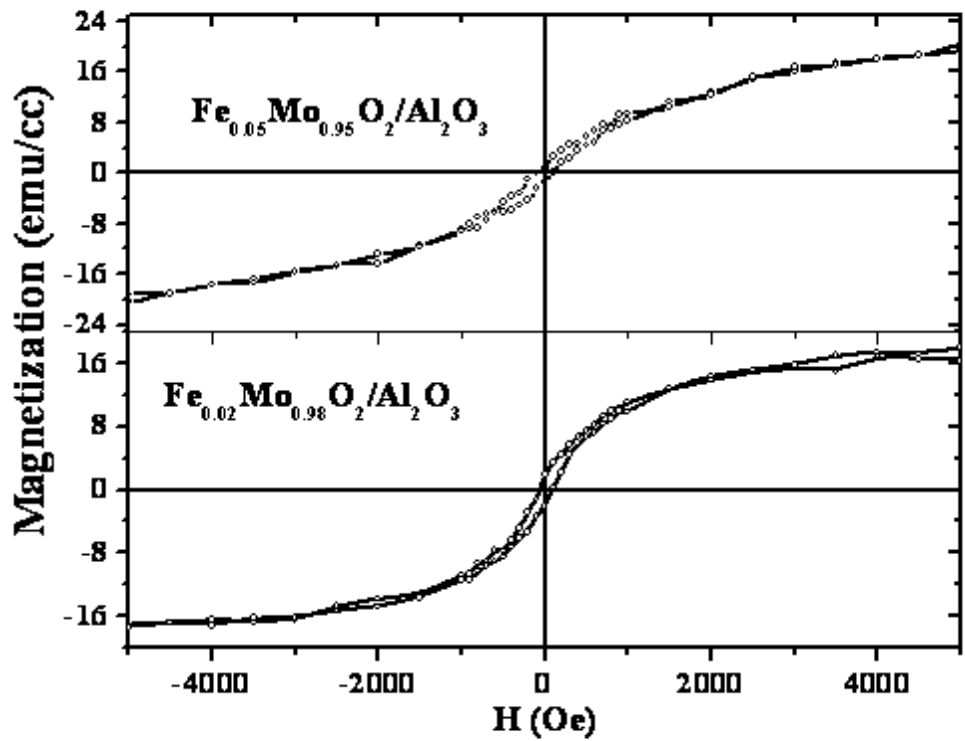